\documentclass[amsmath,amssymb,amsfonts,aps,pre,preprint,superscriptaddress,bibnotes,showpacs,showkeys,longbibliography]{revtex4-1}
\usepackage{epsfig}
\usepackage[english]{babel}
\usepackage{color}

\begin{document}
\title{BLUES function method in computational physics}

\author{Joseph O. Indekeu}
\affiliation{Institute for  Theoretical Physics, KU Leuven, B-3001 Leuven, Belgium}

\author{Kristian K. M\"uller-Nedebock}
\affiliation{Department of  Physics, Stellenbosch University, Stellenbosch, South Africa}

\date{\today}

\begin{abstract}
We introduce a computational method in physics that goes ``beyond linear use of equation superposition" (BLUES). A BLUES function is defined as a solution of a nonlinear differential equation (DE) with a delta source that is at the same time a Green's function for a related linear DE. For an arbitrary source, the BLUES function can be used to construct an exact solution to the nonlinear DE with a different, but related source. Alternatively, the BLUES function can be used to construct an approximate piecewise analytical solution to the nonlinear DE with an arbitrary source. For this alternative use the related linear DE need not be known. The method is illustrated in a few examples using analytical calculations and numerical computations. Areas for further applications are suggested.
\end{abstract}

\maketitle

\section{Setting the stage: An example from population dynamics}
Our aim in this contribution is to provide a method that can assist in constructing a solution to a nonlinear differential equation (DE) with a source term using superposition. In the recent past some efforts in this direction were made, notably by Frasca who developed an extension of the Green's function method in the domain of nonlinear differential equations or nonlinear systems \cite{Frasca}. Here, we propose a different approach. In order to set a concrete stage for illustrating the method we give an example in the field of population dynamics \cite{DMM} and the physics of active matter and recall the Fisher equation \cite{Fisher}, 
\begin{equation}\label{generalfishereq}
\frac{\partial u}{\partial t} = \alpha u(1-u) + D \frac{\partial^2u}{\partial x^2},
\end{equation}
which describes the growth of a (dimensionless) concentration $u(x,t)$  up to a certain level (here normalized to 1) and its diffusion. The growth rate is $\alpha$ and the diffusion constant is $D$. After the transformation: $t \leftarrow \alpha t$, $x \leftarrow \sqrt{\alpha/D} \,x$, we arrive at a dimensionless representation,
\begin{equation}\label{fishereq}
\frac{\partial u}{\partial t} = u(1-u) + \frac{\partial^2u}{\partial x^2}
\end{equation}

The following generalization of the Fisher equation allows for convective motion \cite{Murray}, 
\begin{equation}
\frac{\partial u}{\partial t} + \frac{\partial h(u)}{\partial x} = u(1-u) + \frac{\partial^2u}{\partial x^2}
\end{equation}

For the function $h(u)$ we consider the form $h(u) = a_0 + a_1 u + a_2 u^2$. The first-order term represents a constant convective velocity, which only shifts the propagation velocity. We therefore retain the quadratic term and write $h(u) =  k u^2/2$ \cite{Murray,Mishra2010,Mishra2012}, where $k$ is a positive dimensionless constant. 

It was recently observed that when a delta-source term in the co-moving frame is added,
\begin{equation} \label{Cfishereqdelta}
\frac{\partial u}{\partial t} + k u \frac{\partial u}{\partial x} - u(1-u) - \frac{\partial^2 u}{\partial x^2} =  \frac{1}{k}\delta(x-c(k)\,t),
\end{equation}
where the function $c(k)$ is chosen to match the propagation velocity self-consistently, 
the DE possesses the following traveling wave solution with an exponential tail \cite{IndS},
\begin{equation} \label{solution1}
u(x,t) \equiv U(z) =
\begin{cases}
1, & z<0 \\
e^{-\frac{z}{k}}, & z \geq 0,
\end{cases}
\end{equation}
where $z = x-ct$, with $c$ the velocity of the traveling wave, and $k$ is required to be positive to provide a physical solution. The wave front head is at $z=0$. It is readily verified that (\ref{solution1}) solves \eqref{Cfishereqdelta} with the dispersion relation
\begin{equation}\label{dispersion}
c(k) = k + \frac{1}{k}
\end{equation}
The amplitude of the delta-function source is determined by integrating the DE, expressed in the co-moving frame, from $z= 0^-$ to $z=0^+$ and requiring that the result be an identity. For further convenience, we rewrite the partial DE as an ordinary DE in the variable $z$,
\begin{equation} \label{Cfishereqz}
k(k U -c(k))\frac{\partial U}{\partial z} - kU(1-U) - k\frac{\partial^2 U}{\partial z^2} =  \delta(z),
\end{equation}
and we specify once and for all the boundary conditions that we are interested in, $U(z\rightarrow -\infty) = 1$ and $U(z\rightarrow \infty) = 0$.

Our paper is organized as follows. Section 2 deals with the derivation of a suitable linear DE with a delta source, that is solved by the same piecewise analytic function as the nonlinear DE, which is therefore a Green's function. Section 3 defines the BLUES function and explains how it can be used using superposition of solutions. Section 4 identifies the source which must be combined with the nonlinear DE in order that the piecewise analytic superposition be an exact solution of the nonlinear DE. Section 5 presents an example of a  DE with a gradient-squared nonlinearity for which the same method can be applied. Section 6 gives a perturbation theory viewpoint in which our approach corresponds to the leading (zeroth-order) contribution. This zeroth-order approximation can be used in quite general circumstances. Section 7 illustrates the potential of the method in a few test cases by comparing the analytic zeroth-order solution with the numerical solution. A brief conclusion and outlook on further applications is given in Section 8. 

\section{Deriving a related linear differential equation}
We now attempt to linearize the DE in $U$ about $U=1$. We write $U= 1 +\epsilon$ and obtain the following linear DE in zeroth order (i.e., neglecting $\epsilon$ but not its derivatives),
\begin{equation} \label{Cfishereqlinearz}
k(k-c(k))\frac{\partial U}{\partial z} -  k\frac{\partial^2 U}{\partial z^2}=  \delta(z)
\end{equation}
Remarkably, this DE is again solved by \eqref{solution1} with the same dispersion relation \eqref{dispersion}. 

Making use of \eqref{dispersion} we can now define a Green's function $G$ with the help of the linear operator
\begin{equation} \label{linop}
{\cal L}_z \equiv -\frac{\partial }{\partial z} - k \frac{\partial^2 }{\partial z^2},
\end{equation}
so that 
\begin{equation} \label{Green}
{\cal L}_z G(z-z_0) =  \delta(z-z_0)
\end{equation}
Clearly, the Green's function $G(z)$ coincides with \eqref{solution1}.
The solution $U^{(\ell)}_{\phi}$ of the linear DE ${\cal L}_z U^{(\ell)}_{\phi} (z) = \phi(z)$ with an arbitrary source $\phi(z)$ is then given by
\begin{equation} \label{lineararbisource}
U^{(\ell)}_{\phi}(z) =  \int_{-\infty}^{\infty}  \,G(z-z_0) \phi (z_0) dz_0 \equiv (G\ast\phi)(z)
\end{equation}
For example, for a source that is an exponentially decaying ``corner" with decay constant $K$ and amplitude $A$,
\begin{equation} \label{exposource}
\phi (z) = A e^{-\frac{|z|}{K}},
\end{equation}
the solution to the linear DE is
\begin{equation} \label{exposol}
U^{(\ell)}_{\phi}(z) =
\begin{cases}
AK\left (2-\frac{K}{K+k} e^{\frac{z}{K}}\right ), & z<0 \\
\frac{AK}{K-k} \left (K e^{-\frac{z}{K}} -  \frac{2k^2}{K+k} e^{-\frac{z}{k}}\right ), & z \geq 0,
\end{cases}
\end{equation}
This solution, as well as its first and second derivatives, are continous in $z=0$. The third derivative jumps from $-AK^{-1}(K+k)^{-1}$ at $z=0^-$ to $ A(2K+k)(Kk)^{-1}(K+k)^{-1}$ at $z=0^+$. (In general the singularity in the solution is two orders higher than that in the source.) Note that setting $A=(2K)^{-1}$ in \eqref{exposource} allows one to recover the Dirac $\delta$-function source of \eqref{Cfishereqlinearz} in the limit $K \rightarrow 0$. The solution \eqref{exposol} then tends to the solution \eqref{solution1}.

\section{Beyond linear use of equation superposition}
Since \eqref{solution1}  simultaneously solves the linear and the nonlinear DE with the same delta source, we attempt to use its convolution with an arbitrary source beyond the domain of the linear DE and therefore term it a BLUES (Beyond Linear Use of Equation Superposition) function $B(z)$.
We now test to what extent the superposition
\begin{equation} \label{blues}
U^{(\ell)}_{\phi}(z) = \int_{-\infty}^{\infty}  \,B(z-z_0) \phi (z_0) dz_0 \equiv (B\ast\phi)(z)
\end{equation}
can be a useful approximation to the (unknown) solution of the nonlinear DE with the same source function $\phi (z)$,
\begin{equation} \label{nonlineararbisource}
{\cal N}_z U \equiv -c(k)k\frac{\partial U}{\partial z} + k^2 U \frac{\partial U}{\partial z}- k U(1-U) - k\frac{\partial^2 U}{\partial z^2}=  \phi
\end{equation}
Subtracting the nonlinear DE with operator ${\cal N}_z$ from the linear one allows one to define a residual function ${\cal R}_z U(z)$, with ${\cal R}_z$ the residual operator whose action is defined through
\begin{equation} \label{residual}
{\cal R}_z U \equiv {\cal L}_z U - {\cal N}_z U = k(1-U) (k\frac{\partial U}{\partial z} + U)
\end{equation}
Note that for an arbitrary constant trial concentration $U = {\cal O}(1)$ of order unity the residual is of order $k$. 
As long as ${\cal R}_z U(z)$ is everywhere small (compared to $k$) the error involved in replacing the true solution by the approximation $U^{(\ell)}_{\phi}(z)$ can considered to be small. Note that the residual is zero in the BLUES function, ${\cal R}_z B=0$. 

Clearly, the residual must vanish at the boundaries in order for the solution to satisfy the boundary conditions. This implies that, by taking $z \rightarrow - \infty$ in \eqref{blues}, that the source cannot be arbitrary, but must be normalized as follows, 
\begin{equation} \label{bluesnorm}
\int_{-\infty}^{\infty}  \, \phi (z) dz = 1,
\end{equation}
in order for the superposition \eqref{blues} to be of any use. It is conspicious from \eqref{exposol} that for the proper choice $A=(2K)^{-1}$ in \eqref{exposource} the residual is small for small $K$. Therefore our method allows one to obtain a good approximation to the solution of the nonlinear DE with a strongly localized, but not necessarily infinitely sharply localized, source. Let us quantify this further.

The residual function for the normalized exponential corner source (with $A=(2K)^{-1}$ )
is found to be
\begin{equation} \label{residualf}
{\cal R}_z U^{(\ell)}_{\phi}(z) =
\begin{cases}
\frac{Kk}{2(K+k)} e^{\frac{z}{K}}\left ( 1-\frac{1}{2}e^{\frac{z}{K}}\right ), & z<0 \\
\frac{k}{2}\left (1- \frac{1}{K-k} \left (\frac{K}{2} e^{-\frac{z}{K} } -  \frac{k^2}{K+k} e^{-\frac{z}{k}} \right )\right ) e^{-\frac{z}{K}}, & z \geq 0,
\end{cases}
\end{equation}
Inspection of this function for small $K$ shows that its magnitude is of order $K$, so that the approximation is good for $K \ll k$.

A measure of the significance of the error can be obtained by integrating the residual over space. This leads to the residual functional $\hat {\cal R}[U]$. Since the integral of the exponential tail in \eqref{solution1} is $k$, we can consider $\hat {\cal R}[U]$ to be small when its value is small compared to $k^2$. For our example we obtain the value
\begin{equation} \label{residualF}
\hat{\cal R}[U^{(\ell)}_{\phi}]= \frac{K^2k(3K+4k)}{4(K+k)^2}
\end{equation}
For small $K$ (strongly localized source) this is of order $K^2$, so that the error is of order $(K/k)^2$ compared to unity. 

\section{Constructing the exact source}
We now move on to a different perspective, which is a non-perturbative one. Inspection shows that the residual function can be used to redefine the source  so that the solution $U^{(\ell)}_{\phi}(z)$ of the linear DE with source $\phi(z)$ solves the nonlinear DE with source $\psi (z) \equiv \phi (z) - R_z U^{(\ell)}_{\phi}(z)$. This correspondence can be written in the form
\begin{equation} \label{nonlinearcorresp}
U^{(\ell)}_{\phi}(z) = U^{(non\ell)}_{\psi}(z),
\end{equation}
with $U^{(non\ell)}$ the solution to the nonlinear DE.
We conclude that solutions of the linear DE with an arbitrary source, constructed using the Green's function method, also solve the nonlinear DE exactly for a different, but related, source. The newly constructed source makes up for the drawback that a nonlinear DE does not allow one to apply superposition of solutions. To distinguish the two sources, we denote the function $\phi$ henceforth by {\em resource} and the function $\psi$ by source. The resource is used to construct both the solution (by superposition) and the source (by using the residual operator) of the nonlinear DE.

In this approach it is neither necessary that the source be strongly localized, nor that the residual function be small. We illustrate this with the following example. We let the decay constant $K$ of the source function \eqref{exposource} tend to the decay constant $k$ of the traveling wavefront, and choose $A=(2k)^{-1}$, so that the resource function is
\begin{equation} \label{resource}
 \phi (z)=
\begin{cases}
\frac{1}{2k}  \,e^{\frac{z}{k}}, & z<0 \\
\frac{1}{2k} \,e^{-\frac{z}{k} }, & z \geq 0
\end{cases}
\end{equation}
This simplifies our system to a one-parameter problem with a resource that is localized about $z=0$ and spread out to the same extent as the wavefront. Using \eqref{exposol} and taking the limit $K \rightarrow k$ we obtain the following exact solution 
\begin{equation} \label{exposolKisk}
U^{(\ell)}_{\phi}(z) = U^{(non\ell)}_{\psi}(z)=
\begin{cases}
1-\frac{1}{4} e^{\frac{z}{k}}, & z<0 \\
\left (\frac{3}{4} + \frac{z}{2k} \right ) e^{-\frac{z}{k}}, & z \geq 0,
\end{cases}
\end{equation}
and the exact source (with respect to the nonlinear DE) in the limit $K \rightarrow k$ satisfies
\begin{equation} \label{reconsoKisk}
\psi (z) =\phi (z)- {\cal R}_z U^{(\ell)}_{\phi}(z) =
\begin{cases}
\left (\frac{1}{2k}- \frac{k}{4}\right )  e^{\frac{z}{k}} + \frac{k}{8}e^{\frac{2z}{k}}, & z<0 \\
\left (\frac{1}{2k}- \frac{k}{2}\right ) e^{-\frac{z}{k} } + \left (\frac{3k}{8} + \frac{z}{4} \right ) e^{-\frac{2z}{k}}, & z \geq 0,
\end{cases}
\end{equation}
which is well localized about $z=0$. Note that some restrictions on $k$ are in order to ensure positivity of the source. 

\section{An example with a gradient-squared nonlinearity}
In order to diversify somewhat our cases we present an example with a gradient-squared nonlinearity, very common in physics problems. The nonlinear partial DE with delta source,
\begin{equation}\label{eq-m=2}
\frac{\partial u}{\partial t}  - u(1-u) - \frac{k^2}{2} \left(\frac{\partial u}{\partial x}\right)^2 - (1 + \frac{k^2}{2} u) \frac{\partial^2 u}{\partial x^2} =  (\frac{k}{2} +\frac{1}{k}) \,\delta(x-c(k)\,t),
\end{equation}
contains a gradient-squared term and also possesses an exact wavefront solution corresponding to the simple sharp tail \eqref{solution1} and again with the same velocity dispersion \eqref{dispersion}.
We recast this partial DE into the ordinary form
\begin{equation}\label{eq-m=2ordinary}
-(k+\frac{1}{k}) \frac{\partial U}{\partial z}  - U(1-U) - \frac{k^2}{2} \left(\frac{\partial U}{\partial z}\right)^2 - (1 + \frac{k^2}{2} U) \frac{\partial^2 U}{\partial z^2} =  (\frac{k}{2} +\frac{1}{k}) \,\delta(z),
\end{equation}
and point out that the related linear DE, obtained by replacing $\left(\frac{\partial U}{\partial z}\right)^2 $ by $U\left(\frac{\partial^2 U}{\partial z^2}\right)$ and then approximating $U$ by unity but keeping all the derivatives intact, 
\begin{equation}\label{eq-m=2ordinarylinear}
-(k+\frac{1}{k}) \frac{\partial U}{\partial z}  - (1 + k^2) \frac{\partial^2 U}{\partial z^2} =  (\frac{k}{2} +\frac{1}{k}) \,\delta(z),
\end{equation}
is solved by the same function \eqref{solution1}, which is therefore a BLUES function. In this case the residual operator is
\begin{equation} \label{secondresidual}
{\cal R}_z U \equiv {\cal L}_z U - {\cal N}_z U =  \frac{2k}{k^2+2}\left (U(1-U)  - k^2 (1-\frac{U}{2})\, \frac{\partial^2 U}{\partial z^2}   + \frac{k^2}{2} \left(\frac{\partial U}{\partial z}\right)^2\right )
\end{equation}
The further calculational development of this example is a simple exercise along the lines of the previous sections.

\section{Perturbation theory viewpoint}
In this section we return to situations in which the source is strongly localized and the residual function is small, so that the solution does not differ much from the solution with a delta source, except in a small neighbourhood of the center of the source. We speculate on a formal perturbation expansion and identify our method as the leading (zeroth-order) implementation of it. 
Assume that the function $B$ solves the nonlinear DE with operator ${\cal N}_z$ and delta source,
\begin{equation} \label{NB}
{\cal N}_z B(z-z_0) = \delta (z-z_0),
\end{equation}
and that $B$ is at the same time a Green's function of the linear DE with operator ${\cal L}_z$,
\begin{equation} \label{LB}
{\cal L}_z B(z-z_0) = \delta (z-z_0),
\end{equation}
then we call $B$ a BLUES function. It has the property that its residual vanishes,
\begin{equation} \label{RB}
{\cal R}_z B(z) = {\cal L}_z B(z) - {\cal N}_z B(z) = 0
\end{equation}

Recalling the solution to the linear DE with an arbitrary source $\phi$,
\begin{equation} \label{GreenL}
U^{(\ell)}_{\phi}(z) =  (B\ast\phi)(z),
\end{equation}
we know from the previous sections that we can view $\phi$ as a resource with respect to the nonlinear DE and that the superposition $U^{(\ell)}_{\phi}(z)$ solves the nonlinear DE with the source
\begin{equation} \label{GreenNL}
{\cal N}_z  (B\ast\phi)(z) = \phi (z) - {\cal R}_z U^{(\ell)}_{\phi}(z)  \equiv \phi - {\cal R} (B\ast\phi) \equiv \psi
\end{equation}
Now we turn the problem around and ask how, for an arbitrary source $\psi (z)$, we can obtain a useful approximation to the function $U^{(non\ell)}_{\psi}(z)$ that solves the nonlinear DE.

Since $B\ast\phi$ solves the nonlinear DE, we attempt to express this function in terms of $\psi$. To this end we write
\begin{equation} \label{invert}
\phi  = \psi  + {\cal R} (B\ast\phi),
\end{equation}
and obtain
\begin{equation} \label{perturbseries}
U^{(non\ell)}_{\psi}(z) = (B\ast\phi)(z) = \int_{-\infty}^{\infty}  dz_0 B(z-z_0) \left ( \psi (z_0)   + {\cal R}_{z_0} (B\ast\phi)(z_0)  \right ) 
\end{equation}
If we were to continue using \eqref{invert} iteratively, we would arrive at the following conjectural expansion in powers of the residual, 
\begin{eqnarray} \label{perturbserieslarge}
&& (B\ast\phi)(z) = \int_{-\infty}^{\infty}  dz_0 B(z-z_0)  \psi (z_0)   + \int_{-\infty}^{\infty}  dz_0 B(z-z_0) {\cal R}_{z_0}  \int_{-\infty}^{\infty}  dz_1 B(z_0-z_1)\psi (z_1)  \nonumber  \\
&+& \int_{-\infty}^{\infty}  dz_0 B(z-z_0) {\cal R}_{z_0}  \int_{-\infty}^{\infty}  dz_1 B(z_0-z_1){\cal R}_{z_1}  \int_{-\infty}^{\infty}  dz_2 B(z_1-z_2)\psi (z_2) + \; ...
\end{eqnarray}
which in compact notation reads,
\begin{equation} \label{perturbseriescompact}
U^{(non\ell)}_{\psi} = B\ast\phi  = \sum_{n=0}^{\infty} B\ast \left [ {\cal R} B\right ]^n \psi
\end{equation}
The first term in this series is the zeroth order term in the residual and is identical to the superposition we have proposed based on the BLUES function. It is useful provided the source is properly normalized. The second term, of first order in the residual, is finite provided we use the properly normalized source. However, the addition of this term is likely to prevent the solution from satisfying the required boundary conditions. Favorable situations arise whenever the BLUES function is itself localized and decays to zero for $|z| \rightarrow \infty$. In such case (assuming the source is localized, too) all the terms in the expansion decay to zero for $|z| \rightarrow \infty$. For more general boundary conditions (such as ours), the perturbation theory viewpoint is still incomplete, but the leading term is well defined. In the next section we test its usefulness.

\section{Comparison with the numerical solution}
 
In this section we compare the approximate (zeroth-order) solution $  B\ast\psi$ with the numerical  solution of the nonlinear DE with some source $\psi$. We consider the nonlinear DE \eqref{nonlineararbisource}, which we rewrite as
\begin{equation} \label{nonlineararbisourceR}
k(k U -k-\frac{1}{k}) \frac{\partial U}{\partial z}-  kU(1-U) - k\frac{\partial^2 U}{\partial z^2}=  \psi
\end{equation}
with BLUES function \eqref{solution1} and we first choose the properly normalized exponential corner source
\begin{equation} \label{exposourceR}
\psi (z) =  \frac{1}{2K}e^{-\frac{|z|}{K}}
\end{equation}
The zeroth-order solution is already given in \eqref{exposol} for arbitrary amplitude $A$, so we obtain
\begin{equation} \label{exposolR}
(B\ast\psi)(z)=
\begin{cases}
1-\frac{K}{2(K+k)} e^{\frac{z}{K}}, & z<0 \\
\frac{1}{2(K-k)} \left (K e^{-\frac{z}{K}} -  \frac{2k^2}{K+k} e^{-\frac{z}{k}}\right ), & z \geq 0
\end{cases}
%\label{eq:analyticapprox1}
\end{equation}
In Figs. 1 and 2 we compare this piecewise analytic approximation (red dashed line) to the numerically exact solution (black solid line) of the nonlinear DE. We also show the BLUES function $B(z)$ (blue dotted line). For $K/k \ll 1$ the approximation accurately follows the numerical solution. Even for $K/k=1/4$ the agreement is still very good, as Fig.1 shows. For larger values of $K/k$ the deviations become apparent, as Fig.2 shows for $K/k = 1/2$. Broadening the source by increasing $K/k$ further causes the numerical solution to change sign at some value of $z$, to become non-monotonic and approach zero from below for large $z$. A source with this broadness is unphysical in our context because the concentration variable $U$ must be positive. On the other hand, the approximation $B\ast\psi$ given in \eqref{exposolR} is by construction always positive. It appears to be a useful approximation in the entire domain of physically acceptable sources. 

Note that for constructing the approximation it is not necessary to know the related linear DE for which $B(z)$ is the Green's function. The quality of the approximation depends on the sharpness of the source (quantified in our example by the number $K/k$). However, if the related linear DE is available, it becomes possible to compare the chosen source to the ``exact" source, which allows one to give an additional visual appreciation of the quality of the approximation.

\begin{figure}[h!]
\centering
\includegraphics[width=0.7\textwidth]{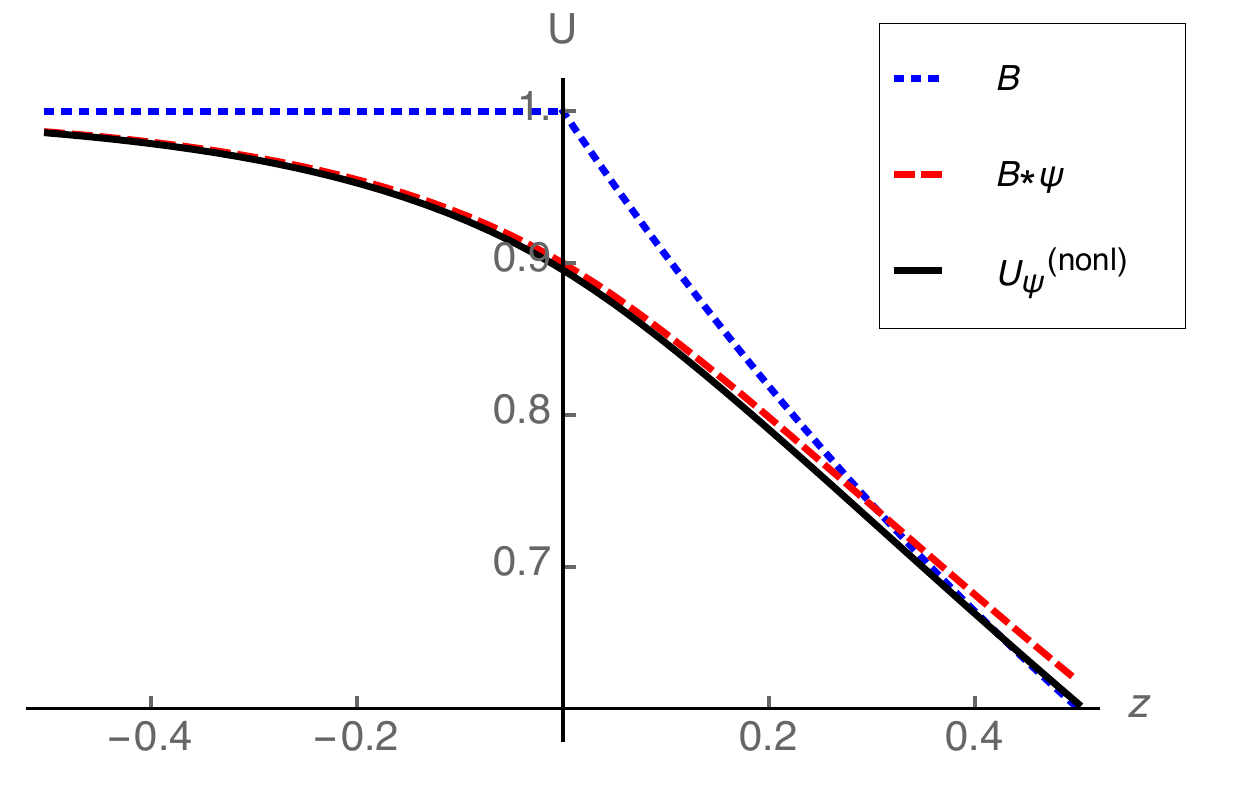}
\caption{Traveling wavefront solution to the nonlinear DE with an exponential corner source. The piecewise analytic approximation (red dashed line, eq.~\eqref{exposolR}) is compared with the numerical solution (black solid line). The BLUES function, which solves the nonlinear DE with a delta source is also shown (blue dotted line). The parameter values are $k=1$ and $K=1/4$.}
\label{1}
\end{figure}
\begin{figure}[h!]
\centering
\includegraphics[width=0.7\textwidth]{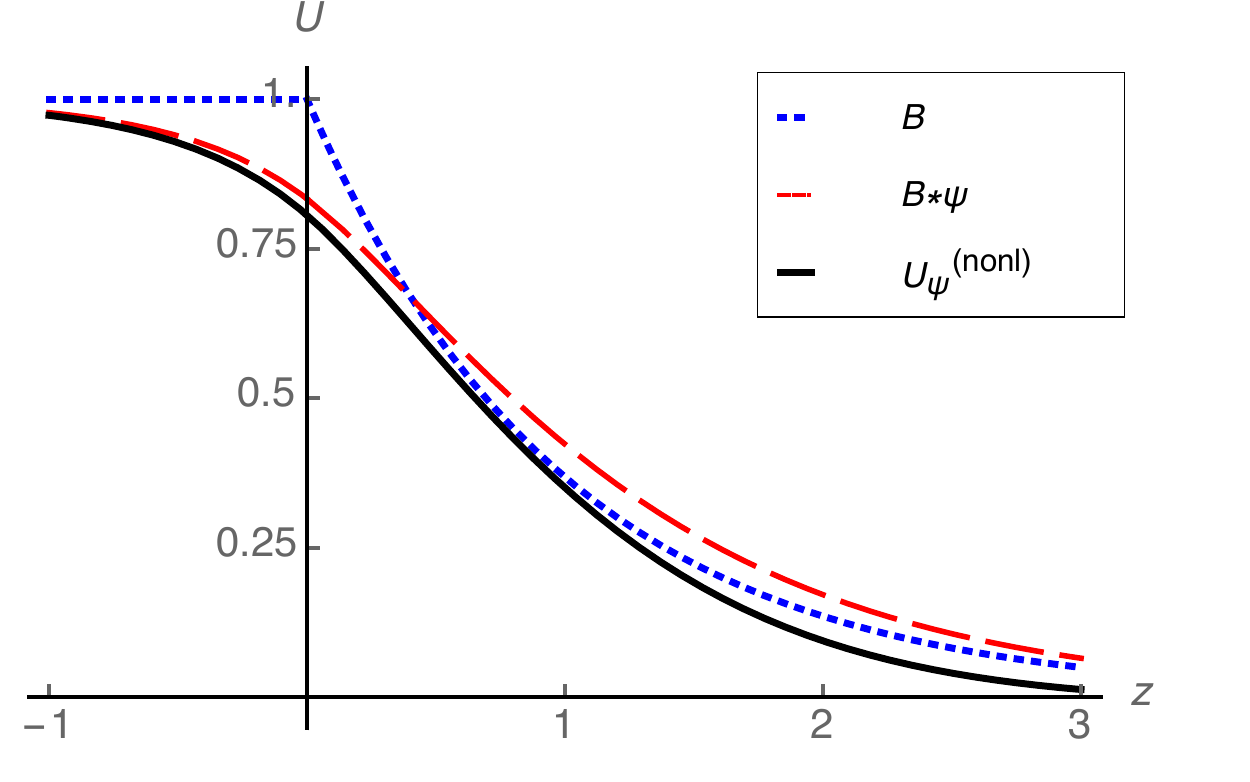}
\caption{Traveling wavefront solution to the nonlinear DE with an exponential corner source. The piecewise analytic approximation (red dashed line, eq.~\eqref{exposolR}) is compared with the numerical solution (black solid line). The BLUES function, which solves the nonlinear DE with a delta source is also shown (blue dotted line). The parameter values are $k=1$ and $K=1/2$.}
\label{2}
\end{figure}

In Figs. 3 and 4 we include the relevant sources together with the piecewise analytic approximation  $ B\ast\psi$. We recall that  $B\ast\psi$ exactly solves the related linear DE 
\begin{equation} \label{lineararbisourceR}
- \frac{\partial U}{\partial z} - k\frac{\partial^2 U}{\partial z^2}=  \psi,
\end{equation}
while the same function $ B\ast\psi$ exactly solves the nonlinear DE \eqref{nonlineararbisourceR} with the different, but related source $\chi$, which follows from subtracting \eqref{residualf} from \eqref{exposourceR},
\begin{equation} \label{residualfR}
\chi (z)= \psi(z) - {\cal R}_z (B\ast\phi)(z) =
\begin{cases}
\frac{1}{2K}e^{\frac{z}{K}}-\frac{Kk}{2(K+k)} e^{\frac{z}{K}}\left ( 1-\frac{1}{2}e^{\frac{z}{K}}\right ), & z<0 \\
\frac{1}{2K}e^{-\frac{z}{K}}-\frac{k}{2}\left (1- \frac{1}{K-k} \left (\frac{K}{2} e^{-\frac{z}{K} } -  \frac{k^2}{K+k} e^{-\frac{z}{k}} \right )\right ) e^{-\frac{z}{K}}, & z \geq 0
\end{cases}
\end{equation}
The sources $\psi$ given in \eqref{exposourceR} and $\chi$ given in \eqref{residualfR} are illustrated in Figs. 3 and 4. Clearly, for small $K/k$ (sharp source) the two sources nearly coincide and it is therefore no surprise that the approximation we propose is accurate. Increasing $K/k$ causes the two sources to differ more and this has consequences for the quality of the approximate solution to the nonlinear DE.
\begin{figure}[h!]
\centering
\includegraphics[width=0.7\textwidth]{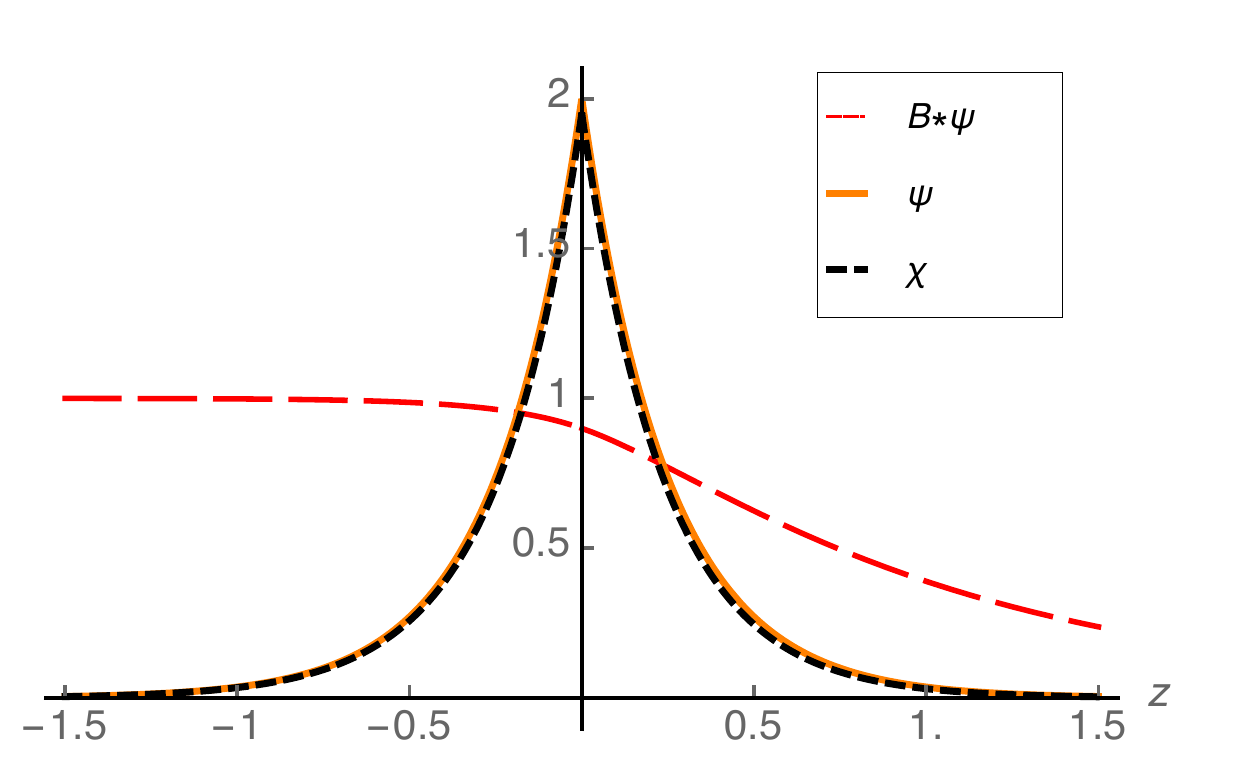}
\caption{The piecewise analytic traveling wavefront solution (red dashed line) solves the linear DE exactly for the exponential corner source (orange solid line) and solves the nonlinear DE exactly for a slightly different, related source (black dashed line). The parameter values are $k=1$ and $K=1/4$. }
\label{3}
\end{figure}
\begin{figure}[h!]
\centering
\includegraphics[width=0.7\textwidth]{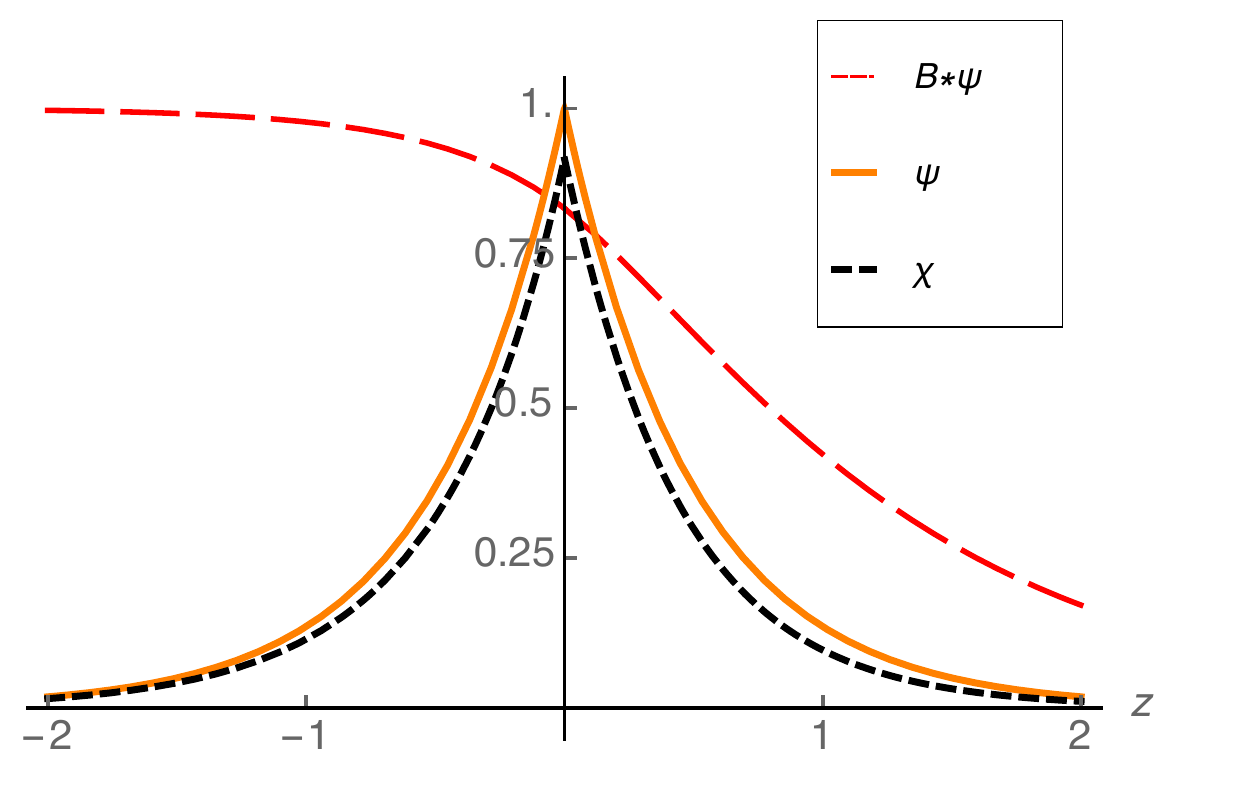}
\caption{The piecewise analytic traveling wavefront solution (red dashed line) solves the linear DE exactly for the exponential corner source (orange solid line) and solves the nonlinear DE exactly for a different, but related source (black dashed line). The parameter values are $k=1$ and $K=1/2$. }
\label{4}
\end{figure}

Next, we demonstrate the method for the same DEs but with a step (piecewise constant) source, properly normalized in accord with \eqref{bluesnorm},
\begin{equation} \label{stepsourceR}
\psi (z) =  
\begin{cases}
\frac{1}{\sigma}, & -\frac{\sigma}{2} \leq z\leq \frac{\sigma}{2} \\
0, & {\rm elsewhere}
\end{cases}
\end{equation}
This choice of source has interesting advantages, as we will illustrate.
The zeroth-order solution is readily calculated,
\begin{equation} \label{stepsolR}
(B\ast\psi)(z) =
\begin{cases}
1,& z < -\frac{\sigma}{2}\\
\frac{k}{\sigma}\left (1 -  e^{-\frac{z+\sigma/2}{k}} \right )+ \frac{1}{\sigma}(\frac{\sigma}{2}-z), & -\frac{\sigma}{2} \leq z \leq \frac{\sigma}{2} \\
\frac{2k}{\sigma} \sinh \frac{\sigma}{2k}  e^{-\frac{z}{k}}, & z > \frac{\sigma}{2}
\end{cases}
\end{equation}
This approximation is particularly useful because outside the support of the source, for $|z| > \sigma/2$, it coincides with the functional form of the exact solution to the nonlinear DE without source, and respects the boundary conditions. Let us clarify this. For $z < -\sigma/2$, $(B\ast\psi)(z)$ equals unity, which trivially solves the nonlinear DE with boundary condition $U(z\rightarrow -\infty) \rightarrow 1$. For $z > \sigma/2$, $(B\ast\psi)(z)$ is of the form $C  \exp(-z/k)  $, which solves the nonlinear DE exactly in this domain. The constant $C>0$ represents the translational invariance of the tail and embodies a shift $s$ through $C =  \exp(s/k)$. The value of $C$ is fixed by the requirement of continuity of the solution at the matching point $z=\sigma/2$ where the nontrivial inner solution in $z \in [-\sigma/2,\sigma/2]$ joins the tail. Clearly, the asymptotic boundary condition $U(z\rightarrow \infty) \rightarrow 0$ is also satisfied.

In Fig. 5 we compare this piecewise analytic approximation (red dashed line) to the numerical solution (black solid line) of the nonlinear DE. We also show the BLUES function $B(z)$ (blue dotted line). Note that for $z> \sigma/2$, $(B\ast\psi)(z)$ is simply a shifted exponential tail with the same decay constant $k$ as the BLUES function. Also the numerical solution is of this simple form in this domain, but with a slightly different shift. For $\sigma/k \ll 1$ the approximation  accurately follows the numerical solution. Even for $\sigma/k=1/2$ the agreement is still very good, as Fig.5 shows. For larger values of $\sigma/k$ the deviations become larger.  For this choice of step source the approximation $B\ast\psi$ given in \eqref{stepsolR} as well as the numerical solution are always positive. 

In order to quantify the difference of the numerical solution and the piecewise analytic one, we can expand both in powers of $\epsilon \equiv (z+\sigma/2)/k$ about the left edge of the step source, $z=-\sigma/2$, and for $\epsilon >0$. Expanding \eqref{stepsolR} to fourth order we obtain
\begin{equation} \label{stepsolRexpanapprox}
(B\ast\psi)(z) =
1-\frac{k}{\sigma}\left (\frac{\epsilon^2}{2} - \frac{\epsilon^3}{3!} + \frac{\epsilon^4}{4!} + {\cal O}(\epsilon^5)\right ), \;\;\;\; \mbox{for} \;  -\frac{\sigma}{2} \leq z \leq \frac{\sigma}{2},
\end{equation}
whereas the readily calculated expansion of the actual solution to the nonlinear DE with step source \eqref{stepsourceR} reads
\begin{equation} \label{stepsolRexpanexact}
U^{(non\ell)}_{\psi} (z) =
1-\frac{k}{\sigma}\left (\frac{\epsilon^2}{2} - \frac{\epsilon^3}{3!} + \frac{(1+k^2)\epsilon^4}{4!} + {\cal O}(\epsilon^5)\right ), \;\;\;\; \mbox{for} \;  -\frac{\sigma}{2} \leq z \leq \frac{\sigma}{2}
\end{equation}
Clearly, the difference starts to appear only from the fourth-order term and is such that the actual solution lies slightly below the piecewise analytic approximation. 

Likewise, we can expand in powers of $\mu \equiv (z-\sigma/2)/k$ about the right edge of the step source, at $z=\sigma/2$, and for $\mu >0$. We again carry the expansions through to four relevant terms. Expanding \eqref{stepsolR} to third order we obtain
\begin{equation} \label{stepsolRexpanapproxR}
(B\ast\psi)(z) =
\frac{k}{\sigma}\left ( (1-e^{-\frac{\sigma}{k}}) +  (e^{-\frac{\sigma}{k}}-1) \mu -  e^{-\frac{\sigma}{k}}\frac{\mu^2}{2} +  e^{-\frac{\sigma}{k} } \frac{\mu^3}{3!} +  {\cal O}(\mu^4)\right ), \;\;\;\; \mbox{for} \;  -\frac{\sigma}{2} \leq z \leq \frac{\sigma}{2},
\end{equation}
whereas the readily calculated expansion of the actual solution to the nonlinear DE with step source \eqref{stepsourceR} reads, with the notation $U^{+} \equiv U^{(non\ell)}_{\psi} (\sigma/2)$,
\begin{equation} \label{stepsolRexpanexactR}
U^{(non\ell)}_{\psi} (z) =
U^+ -U^+ \mu + (U^+ - \frac{k}{\sigma}) \frac{\mu^2}{2} + \left (\frac{k}{\sigma}(1+k^2) -(1+\frac{k^3}{\sigma} ) U^+ \right )  \frac{\mu^3}{3!} + {\cal O}(\mu^4),  \mbox{for} \;  -\frac{\sigma}{2} \leq z \leq \frac{\sigma}{2}
\end{equation}
The value of $U^+$ can be calculated perturbatively by requiring that, in each order of the expansion, $U^{(non\ell)}_{\psi} (-\sigma/2) =1$. This leads to the following sequence of approximations,
\begin{equation} \label{first}
U^+ = (1+\frac{\sigma}{k})^{-1},  \;\;\;  \mbox{to first order in $\mu$}, 
\end{equation}
\begin{equation} \label{second}
U^+ = (1+\frac{\sigma}{2k}) (1+\frac{\sigma}{k}+ \frac{\sigma^2}{2k^2})^{-1},  \;\;\;  \mbox{to second order in $\mu$}, 
\end{equation}
and
\begin{equation} \label{third}
U^+ = \left (1+\frac{\sigma}{2k}-(1+k^2)\frac{\sigma^2}{6k^2} \right) \left (1+\frac{\sigma}{k}+ \frac{\sigma^2}{2k^2}-(1+\frac{k^3}{\sigma})\frac{\sigma^3}{6k^3}\right)^{-1},  \;\;\;  \mbox{to third order in $\mu$} 
\end{equation}
Incidentally, the fourth-order result is less practical because it involves a quadratic equation in $U^+$. Inspection of the expansions \eqref{stepsolRexpanapproxR} and \eqref{stepsolRexpanexactR} shows that i) the value of $U^+$ is in general different from the approximation $\frac{k}{\sigma} (1-e^{-\frac{\sigma}{k}})$, and ii) the expansions are formally the same up to and including the term of second order in $\mu$ and would be identical to that order if we would work with equal values for the leading term $U^+$. 

\begin{figure}[h!]
\centering
\includegraphics[width=0.7\textwidth]{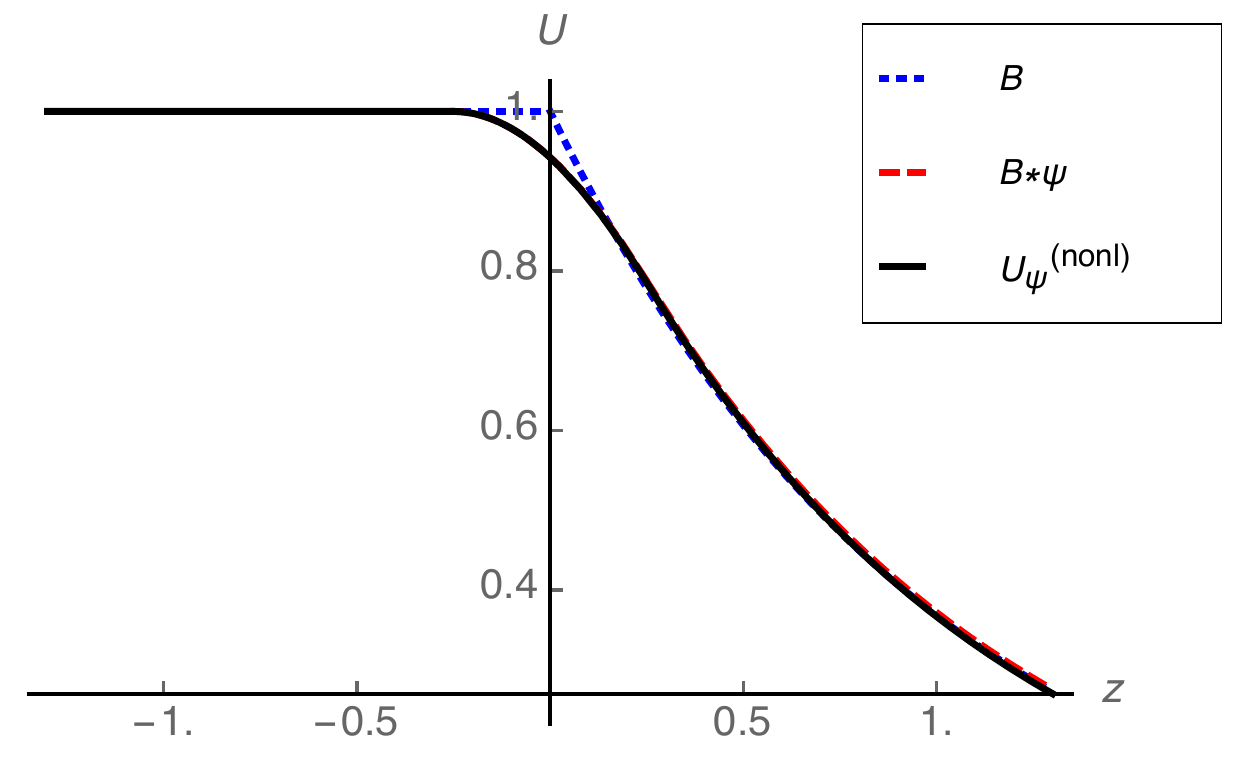}
\caption{Traveling wavefront solution to the nonlinear DE with a step source. The piecewise analytic approximation (red dashed line, eq.~\eqref{stepsolR}) is compared with the numerical solution (black solid line). The numerical solution and analytical zeroth-order approximation are indistinguishable. The BLUES function, which solves the nonlinear DE with a delta source is also shown (blue dotted line). The parameter values are $k=1$ and $\sigma=1/2$.}
\label{5}
\end{figure}

We now return to the properties of the approximate solution \eqref{stepsolR}. The residual of this approximate solution  is readily calculated and is everywhere positive (which easily follows using the inequality $\exp(1-x) \geq 1-x$).
Next, we include the relevant sources  in the figures and recall that the piecewise analytic function $ B\ast\psi$ exactly solves the related linear DE  \eqref{lineararbisourceR}
while the same function $B\ast\psi$ exactly solves the nonlinear DE \eqref{nonlineararbisourceR} with a different, but related source $\chi$, which is readily calculated, 
\begin{equation} \label{residualfRstep}
\chi (z)= \psi (z)  -{\cal R}_z (B\ast\psi)(z) =
\begin{cases}
\frac{1}{\sigma}-\frac{k}{\sigma} (\frac{\sigma}{2} -z)  \left ( \frac{1}{\sigma} (\frac{\sigma}{2} +z)- 
\frac{1}{\sigma} [ 1 - e^{-\frac{z + \sigma/2}{k }}] \right  ), & -\frac{\sigma}{2} \leq z \leq \frac{\sigma}{2} \\
0, & {\rm elsewhere}
\end{cases}
\end{equation}
The sources $\psi$ given in \eqref{stepsourceR} and $\chi$ given in \eqref{residualfRstep} are illustrated in Figs. 6 and 7 for two different values of $\sigma$ and for $k=1$. Clearly, for small $\sigma/k$ (sharp source) the two sources nearly coincide and our approximation is accurate. Increasing $\sigma/k$ causes the two sources to differ more.

\begin{figure}[h!]
\centering
\includegraphics[width=0.7\textwidth]{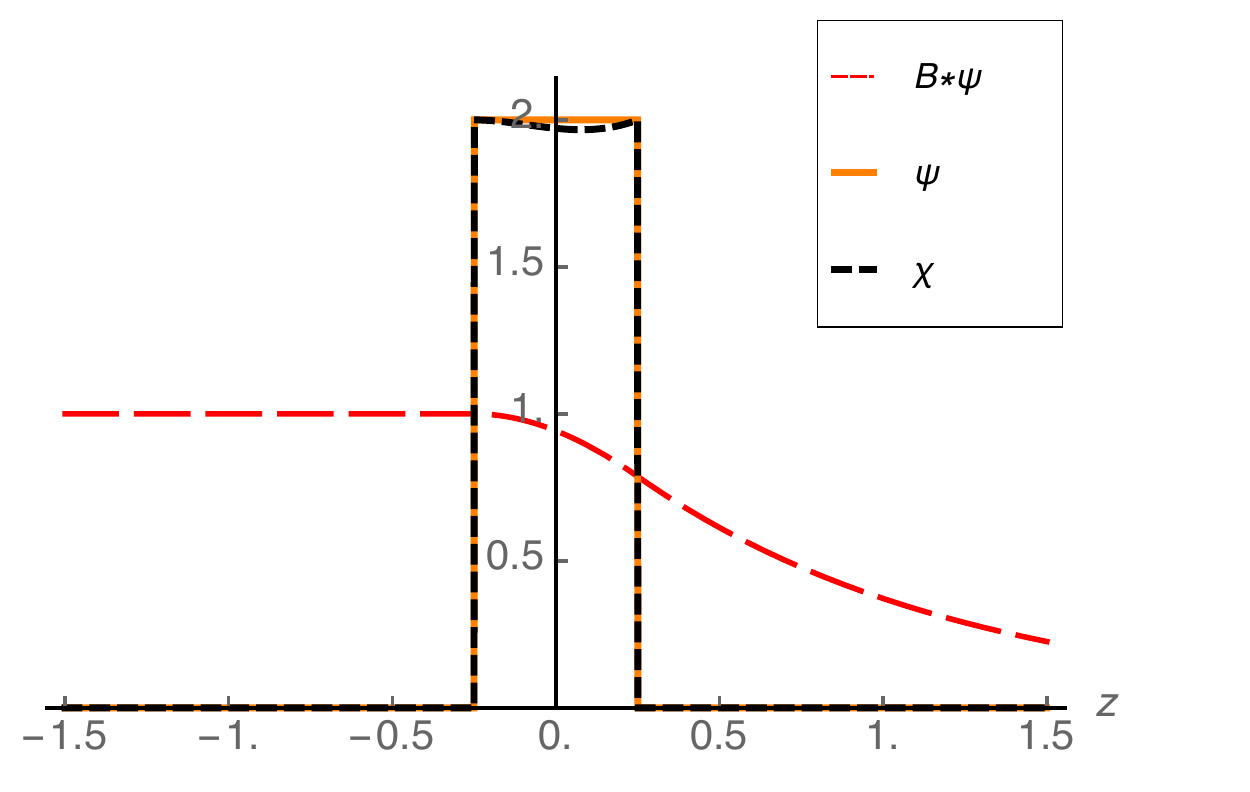}
\caption{The piecewise analytic traveling wavefront solution (red dashed line) solves the linear DE exactly for the step source (orange solid line) and solves the nonlinear DE exactly for a slightly different, related source (black dashed line). The parameter values are $k=1$ and $\sigma =1/2$.}
\label{6}
\end{figure}
\begin{figure}[h!]
\centering
\includegraphics[width=0.7\textwidth]{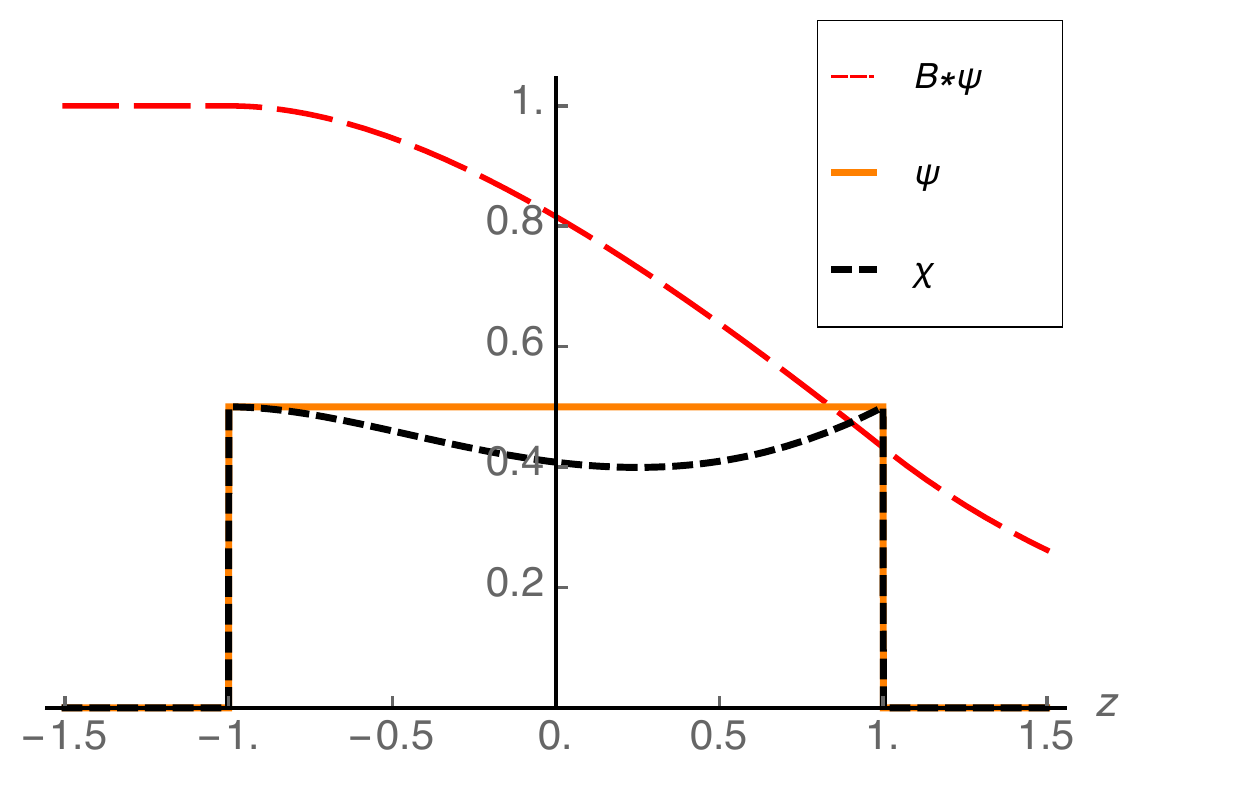}
\caption{The piecewise analytic traveling wavefront solution (red dashed line) solves the linear DE exactly for the step source (orange solid line) and solves the nonlinear DE exactly for a different, but related source (black dashed line).  The parameter values are $k=1$ and $\sigma=2$. }
\label{7}
\end{figure}

\section{Conclusion and outlook}
We conclude that an interesting computational opportunity arises if one knows the solution $B$ of a nonlinear DE with a delta-function source and if it coincides with the Green's function of a related linear DE. In this case the Green's function superposition that solves the linear DE for an arbitrary source also solves the  nonlinear DE for a different, but related, source. Alternatively, an approximate analytic solution to the nonlinear DE can be constructed to zeroth order in perturbation theory.  In this case the source is arbitrary apart from its normalization, which must be imposed in accord with the boundary conditions. 
This alternative use is quite general, as the related linear DE need not be known. The sharpness of the source is the key factor determining the accuracy of the approximation. Computations to higher order in the perturbation theory appear feasible for cases in which $B$ itself is localized, with sufficiently rapid asymptotic decay.

The usefulness of the method has been illustrated in several examples using step sources or exponential corner sources in the context of nonlinear reaction-diffusion-convection equations. Since we have used superposition of equations involving Green's functions but for nonlinear operators that do not commute with the integration, we have used superposition beyond the linear theory, whence the acronym BLUES (``beyond-linear-use-of-equation-superposition") for the function $B$.

Nonlinear differential equations for which this approach is potentially useful occur in a variety of problems in physics, applied physics and other disciplines. Besides the biophysical example which we discussed, the following come to the speculative mind. In cosmology the field equations in Einstein's theory of general relativity are nonlinear differential equations that relate the metric and curvature of space-time to the material mass that acts as a source term (stress-energy tensor). Traveling wave solutions (gravitational waves) occur naturally in the linearized equations and persist also in the nonlinear equations, needed to describe the powerful cosmic events observed recently by LIGO \cite{LIGO}. In electromagnetism the charge density features as the source term in Maxwell's equations. Nonlinear optics takes into account higher-order terms in electric or magnetic fields necessary for describing accurately the properties of metamaterials with (near-)zero refractive index \cite{Mazur}. For these systems the perturbative point of view of nonlinear optics is currently challenged \cite{Boyd}.

More examples of relevant and active research areas in which the method may be useful, can be given. In particular, our interest also goes to condensed matter systems with propagating liquid or solid fronts, relevant mainly to physics and materials engineering. These systems differ from the active matter example treated in this paper, because in condensed matter there are conservation laws (mainly mass conservation) which constrain to an important extent the form of the applicable differential equations. Also the role of sources and sinks is now different and these concepts are replaced by, for example, force terms that correspond to intermittent pinning of a moving three-phase contact line \cite{review}.

\section*{Acknowledgements} 
JOI thanks Universiteit Stellenbosch for hospitality and KU Leuven for a travel grant, in the frame-work of the bilateral agreement.  This work is based on the research supported in part by the National Research Foundation of South Africa (Grant No.~99116). 

{}
\end{document}